# Comments on "Giant Dielectric Response in the One-Dimensional Charge-Ordered Semiconductor $(NbSe_4)_3I$" and "Colossal Magnetocapacitance and Colossal Magnetoresistance in $HgCr_2S_4$"

Gustau Catalan, James F. Scott,
Department of Earth Sciences, University of Cambridge, Cambridge CB2 3EQ, United Kingdom.

There is much interest on the physics of materials with giant dielectric constant and magnetocapacitance, respectively. Two articles by Loidl and co-workers in PRL this year have made claims about two such systems which we feel merit some further analysis.

The first reports giant dielectric constant and relaxor-like behaviour in one-dimensional semiconductor $(NbSe_4)_3I$ [1]. Unfortunately, such behaviour can also be reproduced qualitatively and quantitatively by assuming that the dielectric has regions of different conductivity[3]. In this classic Maxwell-Wagner scenario we have made a simple calculation assuming that the two types of semiconducting region have the same resistivity at room temperature as reported by the authors (1.6 Ωcm), and we have also assumed that the geometry is such that there are wide regions whose conductivity has the measured activation energy (0.13 eV) [1] separated by very narrow regions (thickness ratio $t_i/t=0.001$) with a bigger gap of ~0.5 eV, typical for traps in semiconductors. The intrinsic dielectric constant is the reported value for high-frequency (~100) [1]. With these minimal assumptions, the real part of the dielectric constant has been calculated, and is shown in figure 1. The quantitative and qualitative similarities to the results reported in [1] are evident.

Starešinić *et al.* tried different electrodes and capacitor geometries, always with the same results below 200-250K (above this temperature they saw variations in dielectric constant which were correctly attributed to interface effects). Since the samples are also single crystal (and therefore free of grain boundaries), interfacial effects in the temperature range of interest were ruled out, and so one may raise the question of what is the origin of the heterogeneous conductivity. We do not have a conclusive answer to this, but we note that this being a 1-dimensional semiconductor, any breaks in the conduction chains (due to impurities, micro-cracks, phase boundaries, etc.) can present a bigger conduction gap than that inside the chains. This model is tentative, but it does not require any new physics to fully describe all the reported experimental results, so it should be given full consideration before attempting more sophisticated explanations.

The second work that we would like to comment on is the report of colossal magnetocapacitance and colossal magnetoresistance in the compound $HgCr_2S_4$, another semiconducting dielectric[2]. It was pointed out by one of us [4] that colossal magnetocapacitance could indeed be achieved in magnetoresistive materials that were not actually magnetoelectric. This has been noted by Weber *et al*. [2], albeit with the caution that again interfacial (contact) effects seem to be precluded by the fact that the use of different types of contact did not substantially change their results, so the source of heterogeneous conductivity cannot be a contact-related depletion layer. We agree with that view, but we note that single crystal samples of $HgCr_2S_4$ grown using a different method (without fluorine flux) by Prof. Sang-Wook Cheong of Rutgers University show no dielectric anomalies [7], so it may be that a fluorine-doped surface layer is present in the samples analyzed in [2]. We note also that an intrinsic origin of the dielectric anomaly is ruled out by recent ab-initio calculations which show no phonon softening at any temperature for this material [8]. In sum, while understanding the origin of the large magnetoresistance it is certainly worthy of further work, the clues available so far point to the magnetocapacitance being just a consequence of such magnetoresistance.

The main idea that we want to convey here is that giant dielectric constants and other exotic dielectric phenomena in dielectrics that are not good insulators (such as the ones analyzed here, but also, famously, $CaCu_3Ti_4O_{12}$ [9]) can be due to conductivity artifacts [3-6, 10]. Dielectric spectroscopy is after all just a form of impedance spectroscopy; when the dielectric loss is as big as reported (values of $tan\delta \sim 1$) resistivity effects are at least as important as dielectric ones, and must be taken into account. If one does so using the simplest possible scheme (the Maxwell-Wagner equivalent circuit) many spectacular dielectric phenomena find rather mundane explanations.

References

1. D. Starešinić, P. Lunkenheimer, J. Hemberger, K. Biljaković, and A. Loidl, Phys. Rev. Lett. **96,** 046402 (2006).
2. S. Weber, P. Lunkenheimer, R. Fichtl, J. Hemberger, V. Tsurkan, and A. Loidl, Phys. Rev. Lett. **96**, 157202 (2006).
3. G. Catalan, D. O'Neill, R. M. Bowman, and J. M. Gregg, Appl. Phys. Lett. **77**, 3078 (2000).
4. G. Catalan, Appl. Phys. Lett. **88**, 102902 (2006).
5. D. O'Neill, R. M. Bownman, and J. M. Gregg, Appl. Phys. Lett. **77**, 1520 (2000).
6. P. Lunkenheimer, V. Bobnar, A. V. Pronin, A. I. Ritus, A. A. Volkov, and A. Loidl, Phys. Rev. B **66**, 052105 (2002).
7. K.M. Rabe, private communication.
8. C. J. Fennie and K. M. Rabe, Phys. Rev. B **72**, 214123 (2005).
9. C. C. Homes, T. Vogt, S. M. Shapiro, S. Wakimoto, and A. P. Ramirez, Science **293**, 673 (2001).
10. D. C. Sinclair, T. B. Adams, F. D. Morrison, and A. R. West, Appl. Phys. Lett. **80**, 2153 (2002).

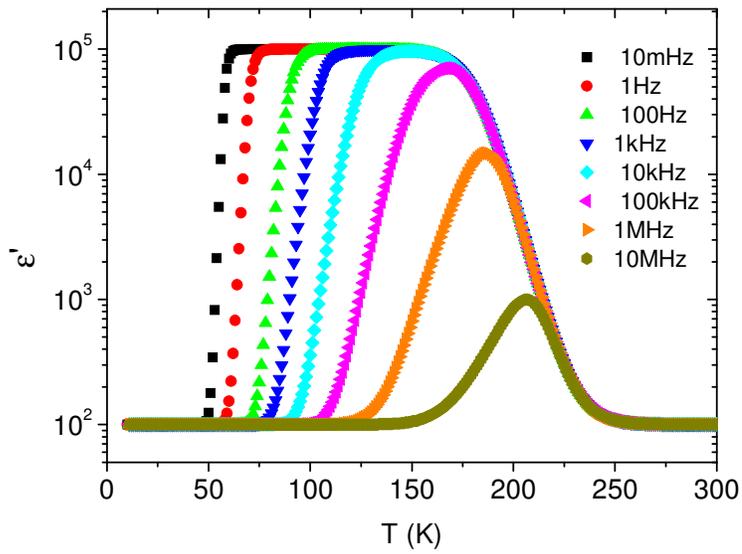

Figure 1: Calculated real part of the dielectric constant in a M-W equivalent circuit with parameters as described in the main text. This calculation does not incorporate the contact-related additional dielectric enhancement above ~200-250K. The similarity with Figure 1 of [1] is nonetheless evident.